\documentclass[pra,twocolumn,aps,letterpaper, preprintnumbers,superscriptaddress]{revtex4}

\usepackage{hyperref}
\usepackage{verbatim}
\usepackage{amsmath}
\usepackage{latexsym}
\usepackage{revsymb}
\usepackage{yfonts}
\usepackage{ifthen}
\usepackage{natbib}
\usepackage{amsfonts}
\usepackage{amsmath}
\usepackage{amssymb}
\usepackage{amsthm}
\usepackage{graphicx}
\usepackage{bm}
\usepackage{bbm}
\usepackage{epsfig,color,amssymb}
\usepackage{subfigure}
\usepackage{amsfonts}
\usepackage{amscd}
\usepackage{amsmath}
\usepackage{multirow}
\usepackage{chemarrow}
\usepackage{dcolumn}
\usepackage{bm}
\usepackage{graphicx}
\usepackage{enumerate}
\usepackage{epsfig}
\usepackage{subfigure}
\usepackage{xcolor}
\usepackage{multirow}
\usepackage{ulem}
\usepackage{braket}
\usepackage{comment}
\usepackage{enumitem}
\usepackage{amsthm}

\renewcommand{\emph}[1]{\textit{#1}}

\begin{document}

\title{Simple security proof of coherent-one-way quantum key distribution}

	\title{Simple security proof of coherent-one-way quantum key distribution}

\author{Rui-Qi Gao}
\affiliation{National Laboratory of Solid State Microstructures, School of Physics and Collaborative Innovation Center of Advanced Microstructures, Nanjing University, Nanjing 210093, China}

\author{Yuan-Mei Xie}
\affiliation{National Laboratory of Solid State Microstructures, School of Physics and Collaborative Innovation Center of Advanced Microstructures, Nanjing University, Nanjing 210093, China}

\author{Jie Gu}
\affiliation{National Laboratory of Solid State Microstructures, School of Physics and Collaborative Innovation Center of Advanced Microstructures, Nanjing University, Nanjing 210093, China}

\author{Wen-Bo Liu}
\affiliation{National Laboratory of Solid State Microstructures, School of Physics and Collaborative Innovation Center of Advanced Microstructures, Nanjing University, Nanjing 210093, China}

\author{Chen-Xun Weng}
\affiliation{National Laboratory of Solid State Microstructures, School of Physics and Collaborative Innovation Center of Advanced Microstructures, Nanjing University, Nanjing 210093, China}

\author{Bing-Hong Li}
\affiliation{National Laboratory of Solid State Microstructures, School of Physics and Collaborative Innovation Center of Advanced Microstructures, Nanjing University, Nanjing 210093, China}

\author{Hua-Lei Yin}
\email{hlyin@nju.edu.cn}
\affiliation{National Laboratory of Solid State Microstructures, School of Physics and Collaborative Innovation Center of Advanced Microstructures, Nanjing University, Nanjing 210093, China}

\author{Zeng-Bing Chen}
\email{zbchen@nju.edu.cn}
\affiliation{National Laboratory of Solid State Microstructures, School of Physics and Collaborative Innovation Center of Advanced Microstructures, Nanjing University, Nanjing 210093, China}

\date{\today}

\begin{abstract}
Coherent-one-way quantum key distribution (COW-QKD), which requires a simple experimental setup and has the ability to withstand photon-number-splitting attacks, has been not only experimentally implemented but also commercially applied. However, recent studies have shown that the current COW-QKD system is insecure and can only distribute secret keys safely within 20 km of the optical fiber length. In this study, we propose a practical implementation of COW-QKD by adding a two-pulse vacuum state as a new decoy sequence. This proposal maintains the original experimental setup as well as the simplicity of its implementation. Utilizing detailed observations on the monitoring line to provide an analytical upper bound on the phase error rate, we provide a high-performance COW-QKD asymptotically secure against coherent attacks. This ensures the availability of COW-QKD within 100 km and establishes theoretical foundations for further applications. 
\end{abstract}

\maketitle

\section{Introduction}\label{sec1}
Quantum key distribution (QKD)~\cite{bennett1984proceedings,ekert1991quantum}, whose security is guaranteed by quantum  laws, allows secret key distribution between two distant parties. Since the Bennett-Brassard 1984 (BB84) protocol~\cite{bennett1984proceedings} was first proposed, numerous QKD schemes have been developed~\cite{scarani2009security,xu2020secure,pirandola2020advances}. To defeat detector attacks~\cite{lydersen2010hacking,xu2020secure}, a viable approach is measurement-device-independent QKD~\cite{lo2012measurement,braunstein2012side}, which has been implemented over a long distance~\cite{zhou2016making,yin2016measurement}. Recently, twin-field QKD~\cite{lucamarini2018overcoming,ma2018phase,wang2018Twin,lin2018simple,yin2019measurement,curty2019simple,cui2019twin,yin2019coherent,minder2019experimental} has also solved this issue and significantly improved the secret key rate.
Another strong restriction in practical QKD is the photon number splitting attack~\cite{brassard2000limitations} on the source side. For example, the coherent-state-based (non-random phase) BB84 protocol can only realize secure key transmissions over 15 km. To overcome this limitation, several approaches such as decoy-state methods~\cite{hwang2003quantum,wang2005beating,lo2005decoy}, strong reference-pulse methods~\cite{koashi2004unconditional}, non-orthogonal coding methods~\cite{scarani2004quantum,tamaki2006unconditionally,yin2016security}, and distributed-phase-reference methods~\cite{inoue2002differential,inoue2003differential,stucki2005fast,sasaki2014practical} have been proposed.

As a type of distributed-phase-reference protocol, coherent-one-way (COW) QKD~\cite{stucki2005fast} has received considerable attention because of its simple and convenient experimental implementation~\cite{stucki2009continuous,stucki2009high,walenta2014fast,korzh2015provably,sibson2017chip,sibson2017integrated,roberts2017modulator,jincheng2020passblock}, which has been deployed in practical quantum communication networks~\cite{peev2009secoqc,Clavis3}.
Considering \emph{the restricted types of collective attacks }~\cite{branciard2008upper}, the key rate depends linearly on the transmittance $\eta$. Additionally, \emph{a variant of  COW-QKD} with a security proof against general attacks was proposed in 2012~\cite{moroder2012security}, and the resulting key rate ``appears to'' be of order $O(\eta^2)$.
To date, all COW-QKD experiments~\cite{walenta2014fast,sibson2017chip,sibson2017integrated,roberts2017modulator,jincheng2020passblock}, including long-distance experiments ~\cite{korzh2015provably,demarco2021realtime}, still employ the original security proof~\cite{branciard2008upper} in which the key rate is of order $O(\eta)$. These experimental results demonstrate the practicability and potential of COW-QKD.

However, the so called ``zero-error attack''~\cite{gonzalez2020upper,trenyi2021zero} involves eavesdropping without breaking the coherence between adjacent non-vacuum pulses (no bit error). Consequently, COW-QKD is insecure if the key rate scales as $O(\eta)$; in fact, the given attack restricts the secure key rate scaling to slightly higher than $\eta^2$. Recently, a novel security proof using semidefinite programming techniques~\cite{characterising2019wang} showed that the transmission distance of COW-QKD using active basis choice is only 20 km. Thus, extending the secure transmission distance of COW-QKD has become an urgent issue.

In this study, we propose a practical implementation of COW-QKD and provide a security proof with precise phase error rate estimation. This proposal keeps all the experimental equipment of the original version unchanged and maintains its ease of implementation. With detailed observations on the monitoring line, we estimate the upper bound on the phase error rate instead of only checking the coherence between adjacent non-vacuum states. We show that the lower bound of the key rate is $0.005\eta^2$, while ensuring the security within 100 kilometers. It is worth mentioning that our secure lower bound for the key rate is approximately 10 times higher than the key rate given by the COW-QKD variant in Ref.~\cite{moroder2012security}, considering that we effectively evaluated the impact of the vacuum states and provided an analytical expression.

\begin{figure*}[t]
	\centering
	\includegraphics[width=13cm]{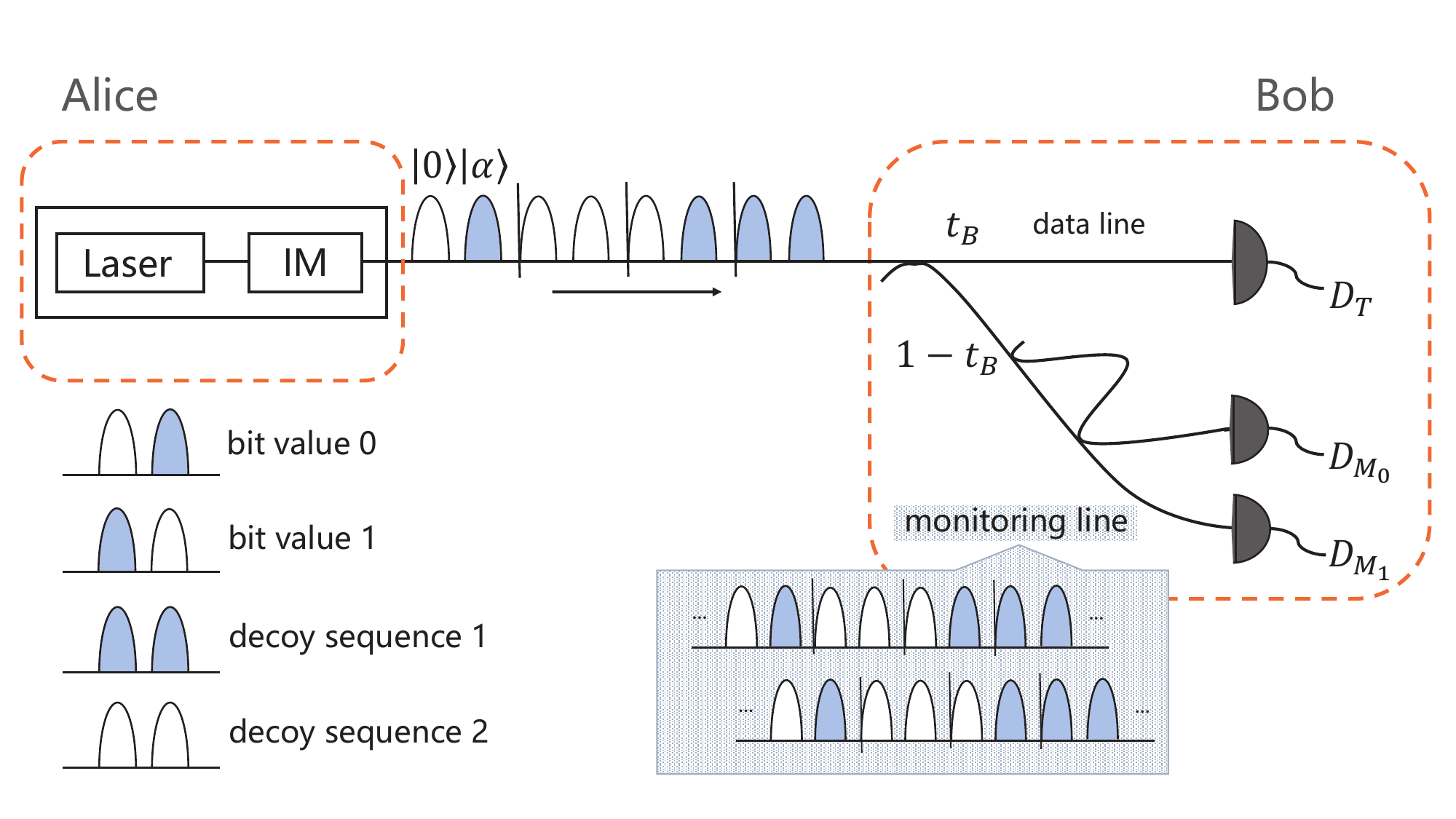}
	\caption{Schematic of  COW-QKD in our work. Alice randomly sends a sequence of states $\ket{0}_{2k-1}\ket{\alpha}_{2k}$, $\ket{\alpha}_{2k-1}\ket{0}_{2k}$,  $\ket{\alpha}_{2k-1}\ket{\alpha}_{2k}$, and $\ket{0}_{2k-1}\ket{0}_{2k}$ to Bob. Then, a beam-splitter of transmittance $t_B$ distributes incoming pulses into the data line and monitoring line at the receiving side, Bob.  The quantum states can be experimentally realized by Alice by modulating $\ket{0}$ or $\ket{\alpha}$ using an intensity modulator (IM) in each time bin. Compared with the original version, where the sequences of states can be prepared in a similar manner, there is no extra technical requirement in our modification. $D_T$, $D_{M_{0}}$, and $D_{M_{1}}$ are the single-photon detectors.}\label{fig1}
\end{figure*}

\section{Protocol description}\label{sec2}
In the original COW protocol, sender Alice encodes logic bits 0 and 1 into a pair of coherent states (non-random phase) $\ket{0_{k}}=\ket{0}_{2k-1}\ket{\alpha}_{2k}$ and $\ket{1_k}=\ket{\alpha}_{2k-1}\ket{0}_{2k}$ at two time points $2k-1$ and $2k$ $(k=1,2,\ldots,K)$,
where $\ket{0}$ is the vacuum state and $\ket{\alpha}$ is a coherent state with mean photon number $\mu=|\alpha|^{2}$. Additionally, Alice sends a two-pulse sequence $\ket{\alpha}_{2k-1}\ket{\alpha}_{2k}$, which is defined as a decoy sequence. Because all non-vacuum pulses share a common phase, Alice can use a mode-locked continuous-wave laser followed by an intensity modulator to prepare weak coherent pulses.

The pulses are then transmitted to the receiver, Bob, through a quantum channel characterized by $\eta$. Bob employs an asymmetric beam-splitter with a transmission coefficient $t_B$ (passive basis choice) to split the incoming pulses into the data line and the monitoring line. We note that Bob can also use an optical switch (active basis choice) instead of the beam-splitter. On the data line, Bob obtains the raw key by measuring the arrival time of each pair of pulses using detector $D_{T}$.
On the monitoring line, Bob checks the coherence between adjacent non-vacuum pulses by observing the measurement outcome of a Mach-Zehnder interferometer with two detectors, $D_{M_0}$ and $D_{M_1}$.
Information leakage can be detected by broken coherence, which can be reflected by the visibility $V=[P\left(D_{M_0}\right)-P\left(D_{M_1}\right)]/[P\left(D_{M_0}\right)+P\left(D_{M_1}\right)]$, where $P\left(D_{M_0}\right)$ [$P\left(D_{M_1}\right)$] is the probability that the detector $D_{M_0}$ ($D_{M_1}$) clicks.
~Alice and Bob also use a random subset of data on the data line to test the bit error rate $E_{\rm b}$. The security proof and secure key rate of  COW-QKD are provided based on these two parameters, $E_{\rm b}$ and $V$~\cite{branciard2008upper}.

Nevertheless, recent studies introduced the so-called zero-error attack~\cite{gonzalez2020upper,trenyi2021zero}. On the one hand, since the signals sent by Alice are linearly independent, Eve can adopt an unambiguous state discrimination measurement to distinguish each signal sent by Alice without introducing error on the data line. On the other hand, the vacuum state in the signals naturally breaks the coherence between adjacent pulses.  Taking advantage of this property, Eve can resend to Bob blocked signal sequences that preserve coherence among consecutive non-empty pulses. Thus, all the security proofs of COW-QKD, which rely on coherence analysis, appear to be unreliable.

As shown in figure~\ref{fig1}, we propose a practical implementation of COW-QKD by adding a two-pulse vacuum state $\ket{0}_{2k-1}\ket{0}_{2k}$ as decoy sequence 2. 
When evaluating the secure key rate, Alice no longer estimates the visibility $V$ to quantify leaked information. Instead, she calculates the gains $Q_{0\alpha}^{M_{i}}$, $Q_{\alpha0}^{M_{i}}$, $Q_{\alpha\alpha}^{M_{i}}$ and $Q_{00}^{M_{i}}$ ($i=0$ or $1$) to estimate the phase error rate, where the superscript $M_i$ represents the clicking detector $D_{M_i}$ on the monitoring line announced by Bob, the subscript refers to the corresponding sequence $\ket{0}_{2k-1}\ket{\alpha}_{2k}$, $\ket{\alpha}_{2k-1}\ket{0}_{2k}$, $\ket{\alpha}_{2k-1}\ket{\alpha}_{2k}$ and $\ket{0}_{2k-1}\ket{0}_{2k}$ sent by Alice. Here, we clarify that if multiple detectors click corresponding to every pair of states, Bob regards this event as one of these detectors clicking randomly \cite{squashing2008beaudry,source2016ma}.

\begin{figure}[t]
	\centering
	\includegraphics[width=8.6cm]{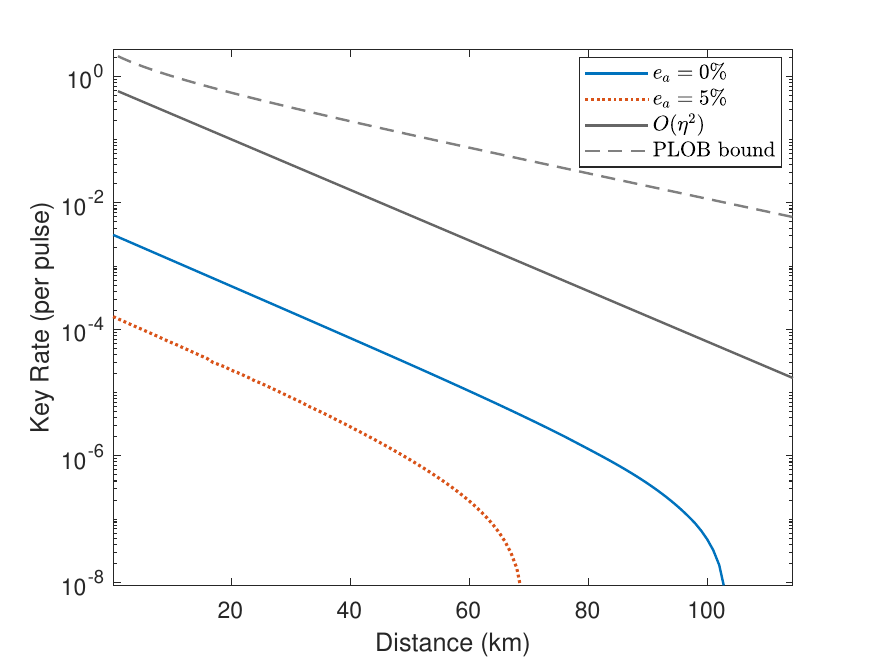}
	\caption{Secret key rates in the asymptotic case using passive basis choice with different misalignment errors, $e_a=0$ and $5\%$. The key rate scales linearly with $0.005\eta^2$ when $e_a=0$, which is much lower than the upper bound on the secret key rate of order $O(\eta^2)$ given in Refs.~\cite{gonzalez2020upper,trenyi2021zero}.}
	\label{fig2}
\end{figure}

\section{Security analysis}
To provide a security proof for COW-QKD, we introduce a virtual entanglement-based protocol as follows:
Here, we redefine the $k$-th optical modes $\ket{0_{z}}=\ket{0}_{2k-1}\ket{\alpha}_{2k}$ and $\ket{1_{z}}=\ket{\alpha}_{2k-1}\ket{0}_{2k}$, where we omit the label $k$ to simplify the presentation.
Let $\ket{0_{x}}=(\ket{0_{z}}+\ket{1_{z}})/\sqrt{N^+}$ and $\ket{1_{x}}=(\ket{0_{z}}-\ket{1_{z}})/\sqrt{N^-}$ be two nonclassical optical modes, where $N^{\pm}=2(1\pm e^{-\mu})$ are the normalization factors. We introduce an ancillary qubit, where $\ket{\pm z}$ and $\ket{\pm x}$ are the eigenstates of the Pauli operators $ Z$ and $ X$ of the qubit, respectively.
Alice prepares $K$ pairs of entangled state
\begin{equation}
	\begin{aligned}
		\ket{\psi}=&\frac{1}{\sqrt{2}}(\ket{+z}_A\ket{0_z}_{A^{\prime}}+\ket{-z}_A\ket{1_z}_{A^{\prime}})\\
		=&\frac{\sqrt{N^+}}{2}\ket{+x}_A\ket{0_{x}}_{A^{\prime}}+\frac{\sqrt{N^-}}{2}\ket{-x}_A\ket{1_{x}}_{A^{\prime}},	
	\end{aligned}
\end{equation}
where the subscript $A$ denotes the ancillary qubit maintained by Alice, and $A^{\prime}$ represents the optical mode sent to Bob. Alice randomly measures the ancillary qubit in the $Z$ or $X$ basis and then acquires the raw keys $\tilde{\textbf{Z}}_{A}$ from the $Z$ basis and $\tilde{\textbf{X}}_{A}$ from the $X$ basis. Then, she sends optical modes to Bob in an insecure quantum channel.
Similar to the practical COW protocol, the optical modes are detected randomly on the data or monitoring line. When observing that the detector $D_{T}$ clicks at the previous moment $\mathcal{T}_{0}$ (the latter moment $\mathcal{T}_{1}$),
Bob records the bit value 0 (1) as the raw key in $\tilde{\textbf{Z}}_{B}$. In addition, the raw key $\tilde{\textbf{X}}_{B}$ is obtained by Bob observing the monitoring line. A detector $D_{M_{0}}$ ($D_{M_{1}}$) click denotes a bit value of 0 (1).

Let $H^{\epsilon}_{\min}\left(\tilde{\textbf{Z}}_A|E\right)$ be the smooth min-entropy characterizing the average probability that Eve guesses $\tilde{\textbf{Z}}_A$ correctly using the optimal strategy with access to the correlations stored in her quantum memory~\cite{konig2009operational}. Let $H^{\epsilon}_{\max}\left(\tilde{\textbf{Z}}_A|\tilde{\textbf{Z}}_B\right)$ be the smooth max-entropy corresponding to the number of extra bits required to reconstruct the value of $\tilde{\textbf{Z}}_A$ using $\tilde{\textbf{Z}}_B$, up to a failure probability of $\epsilon$~\cite{renes2012oneshot}. For later use, we denote the binary Shannon entropy as
$h(a)=-a\log_{2}a-(1-a)\log_{2}(1-a)$, and the size of $\tilde{\textbf{Z}}_{A}$ as $n_{z}$. $\tilde{\textbf{X}}_{A}'$ is the bit string that Alice would have obtained if she had measured in the $X$ basis in virtual protocol, which is actually measured in the $Z$ basis.
Therefore, we have a smooth max-entropy $H^{\epsilon}_{\max}\left(\tilde{\textbf{X}}_{A}'|B\right)\leq n_{z}h(E_{x})$ in the asymptotic limit, where $E_{x}$ is the bit error rate in the $X$ basis. By exploiting the uncertainty relation method for smooth entropies~\cite{tomamichel2011uncertainty}, the asymptotic secure key rate of the virtual entanglement-based protocol can be expressed as
\begin{equation}
	\begin{aligned}
		\tilde{R}&=\frac{1}{\mathcal{P}_{z}K}\left[H^{\epsilon}_{\min}\left(\tilde{\textbf{Z}}_A|E\right)-H^{\epsilon}_{\max}\left(\tilde{\textbf{Z}}_A|\tilde{\textbf{Z}}_B\right)\right]\\
		&=\frac{1}{\mathcal{P}_{z}K}\left[n_{z}-H^{\epsilon}_{\max}\left(\tilde{\textbf{X}}_{A}'|B\right)-n_{z}fh(E_{\rm z})\right]\\
		&=Q_{z}\left[1-h(E_{x})-fh(E_{z})\right],
	\end{aligned}
\end{equation}

\begin{figure}[t]
	\centering
	\includegraphics[width=8.6cm]{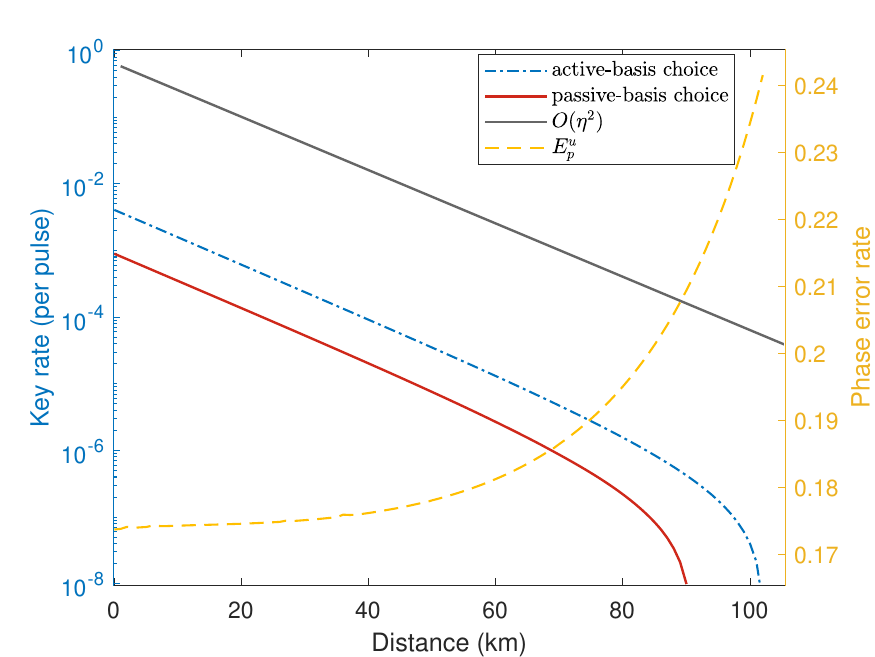}
	\caption{Comparison of secret key rates
		using passive basis choice and active basis choice in the asymptotic case. The dashed yellow line represents the upper bound on the phase error rate $E^u_p$ using active basis choice. The misalignment error $e_a$ is set to $2\%$.}
	\label{fig3}
\end{figure}

where $Q_{z}=n_{z}/(\mathcal{P}_{z}K)=(Q_{0_{z}}^{\mathcal{T}_{0}}+Q_{0_{z}}^{\mathcal{T}_{1}}+Q_{1_{z}}^{\mathcal{T}_{0}}+Q_{1_{z}}^{\mathcal{T}_{1}})/2$ is the gain that Alice measures in the $Z$ basis and Bob detects on the data line, and $\mathcal{P}_{z}$ is the probability that Alice measures in the $Z$ basis and Bob measures on the data line. Here, $f$ denotes the efficiency of error correction and $E_{z}$ is the bit error rate in the $Z$ basis. Moreover, when using the entropic uncertainty relation, the signals detected by Bob’s device should be independent of Alice and Bob’s basis choices \cite{tight2012tomamichel}. This requirement can naturally be satisfied when using active basis choice \cite{moroder2012security,characterising2019wang}, but it needs to be cautiously considered when applying the passive basis choice. Using passive basis choice, Eve can apply classical wavelength attacks \cite{attacking2011li} to partially control Bob’s basis selection and cause weak basis-choice flaws. Nevertheless, the secure key rate decreases only slightly when the wavelength is carefully characterized to minimize this effect \cite{randomness2015li,security2020sun}. Recently, passive basis choice has been widely utilized in various QKD protocols \cite{islam2017provably,security2018rusca,secure2018boaron,boaron2018simple,experimental2019liu,pulsed2020yin,Chen2021integrated}. In this study, using passive basis choice, we assume that the beam-splitter is not controlled by the eavesdropper and use the squashing model \cite{squashing2008beaudry,source2016ma} in the measurement setup.

\begin{figure}[t]
	\centering
	\includegraphics[width=8.6cm]{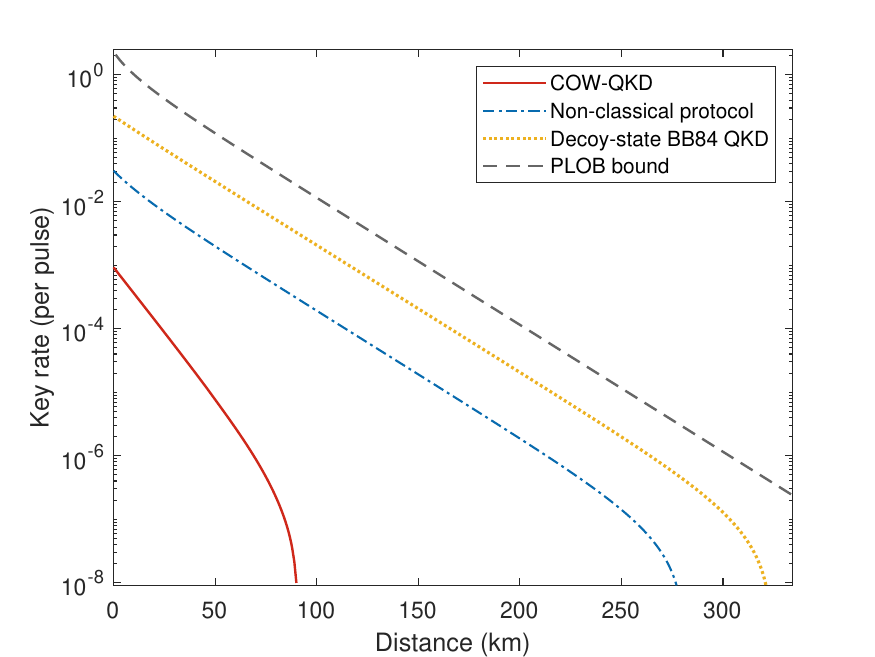}
	\caption{Comparison of asymptotic secret key rates of  COW-QKD in our work, our nonclassical protocol, the decoy-state BB84 QKD and the PLOB bound. The misalignment error $e_a=2\%$ is set as the same. The nonclassical protocol in our work can reach the key rate of order $O(\eta)$.
	}
	\label{fig4}
\end{figure}

In fact, the virtual entanglement-based protocol can be converted to an equivalent prepare-and-measure nonclassical protocol, that is, Alice directly prepares optical modes rather than preparing entangled states and measuring the ancillary qubit system. When Alice chooses the $Z$ basis,  she directly sends the optical modes $\ket{0_{z}}$ and $\ket{1_{z}}$ with probability $1/2$. If Alice selects the $X$ basis, she directly sends nonclassical optical modes $\ket{0_{x}}$ and $\ket{1_{x}}$ with probabilities $N^{+}/4$ and $N^{-}/4$, respectively.

The above facts can be directly seen from the density matrices of the $Z$ and $X$ bases, which are equal in the nonclassical protocol, that is,
\begin{equation}
	\begin{aligned}\label{eq3}
		\rho&=(\ket{0_{z}}\bra{0_{z}}+\ket{1_{z}}\bra{1_{z}})/2\\
		&=(N^{+}\ket{0_{x}}\bra{0_{x}}+N^{-}\ket{1_{x}}\bra{1_{x}})/4.
	\end{aligned}
\end{equation}
Therefore, the bit error rate $E_{z}$ can be obtained directly from the observed gain as follows:
\begin{equation}\label{Ez}
	\begin{aligned}
		E_{z}=\frac{Q_{0_{z}}^{\mathcal{T}_{1}}+Q_{1_{z}}^{\mathcal{T}_{0}}}{Q_{0_{z}}^{\mathcal{T}_{0}}+Q_{0_{z}}^{\mathcal{T}_{1}}+Q_{1_{z}}^{\mathcal{T}_{0}}+Q_{1_{z}}^{\mathcal{T}_{1}}},
	\end{aligned}
\end{equation}
where $Q_{w_{z}}^{\mathcal{T}_{j}}$ represents the gain of the event that Alice sends optical mode $\ket{w_{z}}$ ($w=0$ or $1$) and Bob obtains a click at the $\mathcal{T}_{j}$ ($j=0$ or $1$) moment on the data line, and $Q_{z}=\left(Q_{0_{z}}^{\mathcal{T}_{0}}+Q_{0_{z}}^{\mathcal{T}_{1}}+Q_{1_{z}}^{\mathcal{T}_{0}}+Q_{1_{z}}^{\mathcal{T}_{1}}\right)/2$. Similarly, the bit error rate of the $X$ basis is given by
\begin{equation}\label{EP}
	\begin{aligned}
		E_{x}&=\frac{N^{+}Q_{0_{x}}^{M_{1}}+N^{-}Q_{1_{x}}^{M_{0}}}{N^{+}\left(Q_{0_{x}}^{M_{0}}+Q_{0_{x}}^{M_{1}}\right)+N^{-}\left(Q_{1_{x}}^{M_{0}}+Q_{1_{x}}^{M_{1}}\right)}\\
		&=\frac{N^{+}Q_{0_{x}}^{M_{1}}+\left[2\left(Q_{0_{z}}^{M_{0}}+Q_{1_{z}}^{M_{0}}\right)-N^{+}Q_{0_{x}}^{M_{0}}\right]}{2\left(Q_{0_{z}}^{M_{0}}+Q_{0_{z}}^{M_{1}}+Q_{1_{z}}^{M_{0}}+Q_{1_{z}}^{M_{1}}\right)},
	\end{aligned}
\end{equation}
where $Q_{w_{x(z)}}^{M_{i}}$ represents the gain of the event that Alice senda optical mode $\ket{w_{x(z)}}$ and Bob obtains a click with detector $D_{M_{i}}$ on the monitoring line. In the second equation, we use the relation $N^{+}Q_{0_{x}}^{M_{i}}+N^{-}Q_{1_{x}}^{M_{i}}=2\left(Q_{0_{z}}^{M_{i}}+Q_{1_{z}}^{M_{i}}\right)$, which is obtained using Eqs.~\eqref{eq3}. 

\begin{figure}[t]
	\centering
	\includegraphics[width=8.6cm]{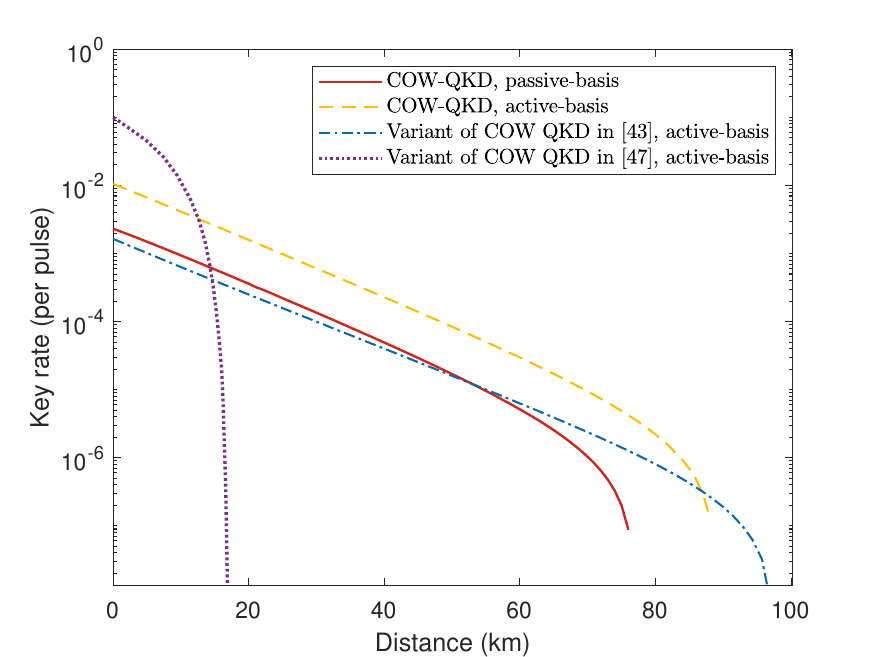}
	\caption{Comparison of asymptotic secret key rates of COW-QKD in this work and the variants of COW-QKD using active basis choice ~\cite{moroder2012security,characterising2019wang}. The misalignment error $e_a$ is set to $1\%$, and the dark count rate is set to $10^{-7}$. The variant in Ref.\cite{moroder2012security} uses 6 optical pulses in each signal block and all blocks of signals share a common phase. The variant in Ref.\cite{characterising2019wang} uses the original setting of the decoy sequence $\ket{\alpha}_{2k-1}\ket{\alpha}_{2k}$. Our work using passive basis choice can still surpass the variant of COW-QKD using active basis choice within 50 km.}
	\label{fig5}
\end{figure}

\begin{table*}
\centering
\begin{tabular}{cccc}\hline\hline
 &Ref.\cite{moroder2012security}&Ref.\cite{characterising2019wang}&This work a.(p.)\\
\hline
Almost maintain the original setting&NO&YES&YES (YES)\\
\hline
Asymptotically secure against coherent attacks&YES&YES&YES (YES)\\
\hline
Secure transmission distance&$\textgreater$90 km&$\textless$20 km&$\textgreater$90 km ($\textgreater$70 km)\\
\hline
\hline
\end{tabular}
\caption{Comparison between variants of the COW-QKD protocol. Here, ``a.'' denotes the active basis choice and ``p.'' denotes the passive basis choice. The dark-count rate $p_d$ is set to $10^{-7}$, and the misalignment error $e_a$ = $1\%$. }
\label{t1}
\end{table*}

Let us now return to the practical COW-QKD protocol. Note that if Alice prepares the encoding sequence $\ket{0}_{2k-1}\ket{\alpha}_{2k}$ or $\ket{\alpha}_{2k-1}\ket{0}_{2k}$ with equal probability, the eavesdropper Eve cannot distinguish this step from the following virtual step: Alice prepares an entangled state $\ket{\psi}$ and measures the ancillary qubit in the $Z$ basis. Consequently, Alice's raw key is identical to $\textbf{Z}_{A}$. The secret key rate in the asymptotic limit can be written as
\begin{equation}
	\begin{aligned}
		R=Q\left[1-h(E_{\rm p}^{\rm u})-fh(E_{\rm b})\right],
	\end{aligned}
\end{equation}
where $Q=Q_{z}$ and $E_{b}=E_{z}$ are the gain and bit error rate, respectively. The phase error rate $E_{\rm p}^{\rm u}$ is the upper bound on the average error probability~\cite{koashi2009simple}, which Bob guesses  as Alice's  bit string  $\textbf{X}_{A}'$ in the virtual entanglement-based protocol. This is equal to the upper bound on the bit error rate of the $X$ basis in the nonclassical protocol. In practice, Alice does not measure qubits in the $X$ basis, which means that one cannot directly acquire the gains $Q_{0_{x}}^{M_{0}}$ and $Q_{0_{x}}^{M_{1}}$ in the nonclassical protocol. However, we can exploit the gains of the other quantum states $\ket{\alpha}_{2k-1}\ket{\alpha}_{2k}$ and $\ket{0}_{2k-1}\ket{0}_{2k}$ which can be directly attained in the COW-QKD protocol to estimate the upper bound $\overline{Q}_{0_{x}}^{M_{1}}$ and the lower bound $\underline{Q}_{0_{x}}^{M_{0}}$. The phase error rate $E_{\rm p}^{\rm u}$ can be expressed as

\begin{equation}\label{up}
	\begin{aligned}
		E_{\rm p}^{\rm u}
		&=\frac{N^{+}\overline{Q}_{0_{x}}^{M_{1}}+\left[2\left(Q_{0_{z}}^{M_{0}}+Q_{1_{z}}^{M_{0}}\right)-N^{+}\underline{Q}_{0_{x}}^{M_{0}}\right]}{2\left(Q_{0_{z}}^{M_{0}}+Q_{0_{z}}^{M_{1}}+Q_{1_{z}}^{M_{0}}+Q_{1_{z}}^{M_{1}}\right)},
	\end{aligned}
\end{equation}
where we use the relation for gain between the nonclassical protocol and the COW-QKD protocol, that is, $Q_{0_{z}}^{M_{i}}=Q_{0\alpha}^{M_{i}}$ and $Q_{1_{z}}^{M_{i}}=Q_{\alpha0}^{M_{i}}$. Under collective attacks, we have the upper bound   $\overline{Q}_{0_{x}}^{M_{1}}$ and lower bound $\underline{Q}^{M_0}_{0_x}$, using the method described in Refs.~\cite{wang2019practical,curty2019simple}. The details can be found in Appendix \ref{Eph}. Using Azuma's inequality~\cite{kazuoki1967weighted}, we remark that the security of COW-QKD can be extended against coherent attacks because the estimation of the phase-error rate will yield consistent results in the asymptotic limit. Security against coherent attacks is presented in Appendix \ref{Coherent attack}.

\section{Numerical simulation}

In our simulation, we assume a dark-count rate of $p_d=10^{-8}$ and detection efficiency of $\eta_d=80\%$. The correction efficiency, $f$, was set to $1.1$. The linear lossy channel is characterized by a transmittance of $\eta=10^{-0.02L}$. The transmission coefficient $t_B$ of the asymmetric beam-splitter is given by the optimization algorithm.

We present the secret key rate of COW-QKD using passive-basis choice with different misalignment errors in figure~\ref{fig2}. Compared with $\eta^2$ and the Pirandola-Laurenza-Ottaviani-Banchi bound (PLOB bound)~\cite{pirandola2009direct,PLOB2017} ($R_{\rm PLOB}=-\log_2(1-\eta)$), the lower bound on the key rate scales is of the order of $O(\eta^2)$,  which is consistent with the result in Ref.\cite{gonzalez2020upper}. The secret key can still be transmitted over 100 km through the optical fiber when misalignment error $e_a=0$.
The secret key rate using active basis choice on the receiving side is also considered. The numerical results of the secret key rate using passive basis choice and active basis choice are presented in figure~\ref{fig3}. Figure~\ref{fig3} also presents the upper bound on the phase error rate $E^u_P$ using active basis choice.

In figure~\ref{fig4}, we show the key rate of the nonclassical protocol using passive basis choice. Comparing the results of the nonclassical protocol with the decoy-state BB84 protocol~\cite{wang2005beating,lo2005decoy} and the PLOB bound~\cite{pirandola2009direct,PLOB2017}, we find that the nonclassical protocol 
also achieves the key rate of order $O(\eta)$.

Additionally, we compare the COW-QKD of this study with the variants of COW-QKD~\cite{moroder2012security,characterising2019wang} in figure~\ref{fig5}. The variant in Ref.\cite{moroder2012security} is considered in the case where all different 3-signal blocks (including 6 optical pulses) share the same phase and Bob applies the active basis choice setup. The variant in Ref.\cite{characterising2019wang} is considered in the case where the decoy sequence keeps the original setting as $\ket{\alpha}_{2k-1}\ket{\alpha}_{2k}$. In figure~\ref{fig5}, the dark count rate $p_d$ is set to $10^{-7}$ and the detection efficiency $\eta_d$ is set to $99\%$. The results reveal that the lower bound on the key rate for the COW-QKD using passive basis choice and active basis choice are both tighter than the result given by the variant in Ref.\cite{moroder2012security}. The secure key distribution in this work also achieves a longer transmittance distance compared with the variant in Ref.\cite{characterising2019wang}. We also almost maintain the original setting of the COW protocol.

Table \ref{t1} summarizes the performance of this study with the variants of the COW-QKD protocol \cite{moroder2012security,characterising2019wang}. The asymptotic security of all these COW-typed QKD protocols has been proven against coherent attacks. Compared with this work, the variant of COW-QKD in Ref. \cite{moroder2012security} also extends the secure key distribution over 90 km using active basis choice; however, it changed the original protocol a lot (coding with m-signal blocks). The variant in Ref. \cite{characterising2019wang} almost maintains the setting of the original COW protocol as in this work. However, its secret keys can only be distributed within 20 km using active basis choice (Here, we consider the variant protocol in Ref.\cite{characterising2019wang}, maintaining the original COW-QKD decoy sequence $\ket{\alpha}_{2k-1}\ket{\alpha}_{2k}$). Our work using passive basis choice also shows good performance, which is presented in  parentheses.

\section{Conclusion}

In this study, we provide an asymptotic security proof for the practical implementation of COW-QKD under coherent attacks and derive the lower bound on the secure key rate. The security proof no longer relies on coherence between adjacent pulses to detect eavesdropping. Instead, we sort the observed quantities to estimate the upper bound on the phase error rate. Our result is tighter than the lower bound on the key rate of the variant COW-QKD \cite{moroder2012security} because we calculate the component of the vacuum states more carefully. Maintaining all current experimental apparatus and techniques, we extend the secure key distribution to 100 km, which paves the way for the secure implementation of COW-QKD.

\acknowledgments
We thank Charles Ci Wen Lim, Nicolas Gisin, and Ignatius William Primaatmaja for enlightening the discussions and making the security proof of this work more rigorous. We thank Amine Iggidr for helping us find 
and correct an incorrect expression $c5$ given in appendix A of the previous version. We gratefully acknowledge support from National Natural Science Foundation of China (No. 61801420), Natural Science Foundation of Jiangsu Province (No. BK20211145), Fundamental Research Funds for the Central Universities (No. 020414380182), Key Research and Development Program of Nanjing Jiangbei New Aera (No. ZDYD20210101), Key-Area Research and Development Program of Guangdong Province (No. 2020B0303040001). 
R.-Q.G. and Y.-M.X. contributed equally to this work.

\appendix

\section{Upper bound on phase error rate}\label{Eph}

Here, we first consider the case in which Bob applies a passive basis choice to distribute incoming pulses into the data and monitoring lines. The gain of the state $\ket{\phi}$  heralded by detector $D_{M_{i}}$ ($i=0$ or $1$) clicking can be given by
\begin{equation}
	Q_{\phi}^{M_{i}}=\bra{\phi}\hat{\mathcal{M}}_{i}^{+}\hat{\mathcal{M}}_{i}\ket{\phi},\label{Mi}
\end{equation}	
where $\hat{\mathcal{M}}_{i}$ is the Kraus operator corresponding to detector $D_{M_{i}}$.

In the nonclassical protocol, the gains of the states $\ket{0_z}$ and $\ket{1_z}$ detected by detector $D_{M_i}$ on the monitoring line, that is, $Q_{0_{z}}^{M_{i}}$ and $Q_{1_{z}}^{M_{i}}$, can be written as
$Q_{0_{z}}^{M_{i}}=Q_{1_{z}}^{M_{i}}=\left(1-p_d\right)^2e^{-t_B\mu\eta}c1\left(1-c1\right)$, where $c1=\left(1-p_d\right)e^{-(1-t_B)\mu\eta/2}$.
The gains $Q_{0_{x}}^{M_{1}}$ and $Q_{0_{x}}^{M_{0}}$ of the nonclassical states $\ket{0_x}$ on the monitoring line are  $Q_{0_{x}}^{M_{0}}=\frac{2}{N^{+}}\left(1-p_d\right)^3\left(1-c1\right)\left[e^{-(1+t_B)\mu/2}c2+e^{-(1-t_B)\mu\eta/2}c3\right]$ and
$Q_{0_{x}}^{M_{1}}=\frac{2}{N^{+}}\left(1-p_d\right)^2c1\{e^{-(1+t_B)\mu/2}\left[c4-\left(1-p_d\right)c2\right]\\
+c3\left(1-c1\right)\}$, where the parameters $c2=e^{(1-t_B)\mu(1-\eta)/2}+e^{-(1-t_B)\mu(1-\eta)/2}$, $c3=e^{-t_{B}\mu\eta}-e^{-t_{B}\mu}$ and $c4=e^{(1-t_B)\mu/2}+e^{-(1-t_B)\mu/2}$.

In the COW protocol, because the nonclassical state $\ket{0_{x}}$ cannot be acquired, we cannot directly estimate the gain $Q_{0_{x}}^{M_i}=\bra{0_x}\hat{\mathcal{M}}_{i}^{+}\hat{\mathcal{M}}_{i}\ket{0_x}$. We express $\ket{0_{x}}$ using the states $\ket{0,0}$, $\ket{\alpha, \alpha}$, and $\ket{\beta,\beta}$ as follows:
\begin{equation}
	\ket{0_{x}}=\frac{e^{-\mu}\ket{0,0}+\ket{\alpha,\alpha}-(1-e^{-\mu})\ket{\beta,\beta}}{e^{-\mu/2}\sqrt{2(1+e^{-\mu})}},\label{NON-CLA}
\end{equation}
where the state $\ket{\beta}=(\ket{\alpha}-e^{-\mu/2}\ket{0})/\sqrt{1-e^{-\mu}}$ denotes the normalized non-vacuum part of the state $\ket{\alpha}$.
The pairs of states $\ket{0,0}$ and $\ket{\alpha,\alpha}$ correspond to decoy sequences sent by Alice. Thus, we can rewrite the gain $Q_{0_{x}}^{M_i}$ as
\begin{equation}
	\bra{0_x}\hat{\mathcal{M}}_{i}^{+}\hat{\mathcal{M}}_{i}\ket{0_x}=\sum_{l,k}s_{l}s_{k}\bra{\phi_{l}}\mathcal{M}_{i}^{+}\mathcal{M}_{i}\ket{\phi_{k}},\label{Qox}
\end{equation}
where, $\ket{\phi_l},\ket{\phi_k}\in\{\ket{0,0},\ket{\alpha,\alpha},\ket{\beta,\beta}\}$, $s_l$, and $s_k$ are the corresponding state coefficients in Eq.~\eqref{NON-CLA}. By utilizing the Cauchy inequality~\cite{wang2019practical,curty2019simple}, we have that
\begin{equation}
	\left|s_{l}s_{k}\bra{\phi_{l}}\mathcal{M}_{i}^{+}\mathcal{M}_{i}\ket{\phi_{k}}\right|
	\le|s_{l}s_{k}|\sqrt{\left|\hat{\mathcal{M}}_{i}\ket{\phi_{l}}\right|^2}\sqrt{\left|\hat{\mathcal{M}}_{i}\ket{\phi_{k}}\right|^2}.
\end{equation} 	
\\
Combining this result with the limit $0\le\bra{\beta,\beta} \hat{\mathcal{M}}_{i}^{+}\hat{\mathcal{M}}_{i}\ket{\beta,\beta} \le 1 $, the upper bound for ${Q}^{M_1}_{0_{x}}$ and the lower bound for ${Q}^{M_0}_{0_{x}}$ can be expressed as
\begin{equation}\label{BOUND}
	\begin{aligned}
		\overline{Q}_{0_{x}}^{M_{1}}=&\frac{1}{N^{+}}\left(e^{\frac{\mu}{2}}\sqrt{Q_{\alpha\alpha}^{M_{1}}}+e^{-\frac{\mu}{2}}\sqrt{Q_{00}^{M_{1}}}\right)^{2}\\
		&+\frac{N^{-}}{N^{+}}\left(\frac{e^{\mu}N^{-}}{4}+e^{\mu}\sqrt{Q_{\alpha\alpha}^{M_{1}}}+\sqrt{Q_{00}^{M_{1}}}\right),\\
		\underline{Q}^{M_0}_{0_{x}}=&\frac{1}{N^{+}}\left(e^{\frac{\mu}{2}}\sqrt{Q_{\alpha\alpha}^{M_{0}}}-e^{-\frac{\mu}{2}}\sqrt{Q_{00}^{M_{0}}}\right)^{2}\\
		&-\frac{N^{-}}{N^{+}}\left(e^{\mu}\sqrt{Q_{\alpha\alpha}^{M_{0}}}+\sqrt{Q_{00}^{M_{0}}}\right).\\
	\end{aligned}
\end{equation}
Here, the gains of state $\ket{\alpha}_{2k-1}\ket{\alpha}_{2k}$ on the monitoring line are $Q_{\alpha\alpha}^{M_{0}}=\left(1-p_d\right)^3\left[1-\left(1-p_d\right)e^{-2\mu(1-t_B)\eta}\right]c5$, 	 and $Q_{\alpha\alpha}^{M_{1}}=p_d\left(1-p_d\right)^3e^{-2\mu(1-t_B)\eta}c5$, where $c5= e^{-2t_B\mu\eta}$ and the gains of the states  $\ket{0}_{2k-1}\ket{0}_{2k}$ on the monitoring line are  $Q_{00}^{M_{i}}=p_d(1-p_d)^3$.

In addition, when Bob applies an active basis choice setup, the gains $Q_{0_{z}}^{M_{i}}$, $Q_{1_{z}}^{M_{i}}$, $Q_{\alpha\alpha}^{M_{i}}$, and $Q_{00}^{M_{i}}$ are given by  $Q_{0_{z}}^{M_{i}}=Q_{1_{z}}^{M_{i}}=c1\left(1-c1\right)$.
$Q_{\alpha\alpha}^{M_{0}}=\left(1-p_d\right)\left[1-\left(1-p_d\right)e^{-2\mu(1-t_B)\eta}\right]$,
$Q_{\alpha\alpha}^{M_{1}}=p_d\left(1-p_d\right)e^{-2\mu(1-t_B)\eta}$ and $Q_{00}^{M_{i}}=p_d(1-p_d)$,
with other formulas unchanged.

\section{Security against coherent attacks}\label{Coherent attack}
Here, we extend the security of COW-QKD against coherent attacks.
We consider a process in which a pair of states is sent at two time points, $2k-1$ and $2k$ $(k=1,2,\ldots,
K)$ as one round, and the entire process of the COW protocol is performed by repeating this round $ K$ times. Following a similar security analysis as in the main text, we assume an equivalent prepare-and-measure nonclassical protocol. Let the Kraus operator $\hat{\mathcal{M}}_{i}^l$ correspond to the announcement of detector $D_{M_{i}}$ associated with the $l$-th round $(l=1,2,\ldots,K)$.
We denote the probability of the event that Alice decides to send the nonclassical optical mode $\ket{0_x}$ as $p_{0_x}$. For any arbitrary small quantity $\epsilon \geq 0$, according to Azuma's inequality~\cite{kazuoki1967weighted}, we have that 
\begin{equation}\label{Azuma2}
    \left| \sum_{l=1}^{K} p_{0_x} Q_{0_x}^{l,{M_i}} - N_{0_x}^{l,{M_i}} \right| < \epsilon \left|K \right|
\end{equation}
holds, except for the minuscule probability, $2e^{-K\epsilon^2/2}$. Here, $Q_{0_x}^{l,{M_i}}$ corresponds to the probability that the nonclassical optical mode $\ket{0_x}$ is detected by the detector $D_{M_i}$ on the monitoring line associated with the $l$-th round. $N_{0_x}^{l,{M_i}}$, as an observed quantity, is the actual number of events that the nonclassical optical mode $\ket{0_x}$ sent by Alice is detected by the detector $D_{M_i}$ on the monitoring line. When $K$ tends to infinity in the asymptotic limit, we can deduce from Eq.~\eqref{Azuma2} that 
\begin{equation}
    \frac{N_{0_x}^{l,{M_i}}}{K} \approx \frac{\sum_{l=1}^{K} p_{0_x} Q_{0_x}^{l,{M_i}} }{K}
\end{equation}
Here, $Q_{0_x}^{l,{M_i}}$ satisfies Eq.~\eqref{BOUND}, and all gains of the COW-QKD protocol in Eq.~\eqref{BOUND} are associated with the $l$-th round (which corresponds to the condition of the security proof against collective attacks). Combining this result with the upper bound on the phase error rate derived from Eq.~\eqref{up} in the main text, we find that the estimation result of the phase error rate is consistent in the asymptotic limit when considering collective and coherent attacks. Thus, the asymptotic security of COW-QKD against coherent attacks is proven.

\end{document}